\PassOptionsToPackage{linesnumbered,ruled,vlined}{algorithm2e}
\documentclass[10pt,conference]{IEEEtran}
\IEEEoverridecommandlockouts

\usepackage{cite}
\usepackage{amsmath,amssymb,amsfonts}
\usepackage{algorithmic}
\usepackage{graphicx}
\usepackage{textcomp}

\usepackage[nolist]{acronym}
\usepackage{float}
\usepackage{fontawesome}
\usepackage{graphicx}
\usepackage{amsmath}
\usepackage{tikz}
\usepackage[ruled,vlined]{algorithm2e}
\usepackage{graphicx}
\usepackage{booktabs} 
\usepackage[linesnumbered,ruled,vlined]{algorithm2e}

\usepackage{graphicx}
\usepackage{subcaption}
\usepackage{url}
\usepackage{booktabs,tabularx}

\usepackage{tabularx}
\usepackage{makecell}

\usepackage{graphicx}

\def\BibTeX{{\rm B\kern-.05em{\sc i\kern-.025em b}\kern-.08em
    T\kern-.1667em\lower.7ex\hbox{E}\kern-.125emX}}
\begin{document}
\acrodef{3GPP}{3rd Generation Partnership Project}
\acrodef{5G}{5th Generation Mobile Network}
\acrodef{6G}{6th Generation Mobile Network}
\acrodef{AI}{Artificial Intelligence}
\acrodef{AI4Net}{\ac{AI} for Networking}
\acrodef{MDP}{Markov Decision Process}
\acrodef{AIDER}{Aerial Image Dataset for Emergency Response}
\acrodef{AMF}{Access and Mobility Management Function}
\acrodef{AIaaS}{Artificial Intelligence-as-a-Service}
\acrodef{AC}{Actor-Critic}
\acrodef{IID}{Independent and Identically Distributed}
\acrodef{B5G}{Beyond Fifth Generation}
\acrodef{BPF}{Berkeley Packet Filter}
\acrodef{CBR}{Constant Bit Rate}
\acrodef{CSV}{Comma-Separated Values}
\acrodef{CPU}{Central Processing Unit}
\acrodef{CNN}{Convolutional Neural Network}
\acrodef{CNNs}{Convolutional Neural Networks}
\acrodef{C-V2X}{Cellular Vehicle to-Everything}
\acrodef{DoS}{Denial of Service}
\acrodef{DDQL}{Double Deep
 Q-learning}
\acrodef{DDoS}{Distributed Denial of Service}
\acrodef{DDPG}{Deep Deterministic Policy Gradient}
\acrodef{DNN}{Deep Neural Network}
\acrodef{DRL}{Deep Reinforcement Learning}
\acrodef{DQN}{Deep Q-Network}
\acrodef{DT}{Decision Tree}
\acrodef{DDQN}{Double Deep Q-Network}
\acrodef{DPDK}{Data Plane Development Kit}
\acrodef{ETSI}{European Telecommunications Standards Institute}
\acrodef{eNWDAF}{Evolved Network Data Analytics Function}
\acrodef{eBPF}{Extended Berkeley Packet Filter}
\acrodef{ECDF}{Empirical Cumulative Distribution Function}
\acrodef{ECDFs}{Empirical Cumulative Distribution Functions}
\acrodef{FIBRE}{Future Internet Brazilian Environment for Experimentation}
\acrodef{GNN}{Graph Neural Networks}
\acrodef{GPU}{Graphics Processing Unit}
\acrodef{GPRS}{General Packet Radio Service}
\acrodef{GTP}{General Packet Radio Service Tunnelling Protocol}
\acrodef{GTP-U}{General Packet Radio Service Tunnelling Protocol User Plane}
\acrodef{HTM}{Hierarchical Temporal Memory}

\acrodef{IAM}{Identity And Access Management}
\acrodef{ICMP}{Internet Control Message Protocol}
\acrodef{IID}{Informally, Identically Distributed}
\acrodef{IoE}{Internet of Everything}
\acrodef{IoT}{Internet of Things}
\acrodef{ITU}{International Telecommunication Union}
\acrodef{IQR}{Interquartile Range}
\acrodef{I/O}{Input/Output}
\acrodef{IP}{Internet Protocol}
\acrodef{KNN}{K-Nearest Neighbors}
\acrodef{KPI}{Key Performance Indicator}
\acrodef{KPIs}{Key Performance Indicators}
\acrodef{LSTM}{Long Short-Term Memory}
\acrodef{LOWESS}{Locally Weighted Scatterplot Smoothing}
\acrodef{MAE}{Mean Absolute Error}
\acrodef{MAD}{Median Absolute Deviation}
\acrodef{ML}{Machine Learning}
\acrodef{MLaaS}{Machine Learning as a Service}
\acrodef{MOS}{Mean Opinion Score}
\acrodef{MAPE}{Mean Absolute Percentage Error}
\acrodef{MSE}{Mean Squared Error}
\acrodef{MEC}{Multi-access Edge Computing}
\acrodef{mMTC}{Massive Machine Type Communications}
\acrodef{MFA}{Multi-factor Authentication}
\acrodef{MLP}{Multi-Layer Perceptron}
\acrodef{MADRL}{Multi-Agent Deep Reinforcement Learning}
\acrodef{MAB}{Multi-Armed Bandit}
\acrodef{MILP}{Mixed Integer Linear Programming}
\acrodef{NWDAF}{Network Data Analytics Function}
\acrodef{Net4AI}{Networking for \ac{AI}}
\acrodef{NS}{Network Slicing}
\acrodef{NFV}{Network Function Virtualization}
\acrodef{NN}{Noisy Neighbor}
\acrodef{NNs}{Noisy Neighbors}
\acrodef{OSM}{Open Source MANO}
\acrodef{OPEX}{Operating Expenditures}
\acrodef{O-RAN}{Open Radio Access Network}
\acrodef{PCA}{Principal Component Analysis}
\acrodef{PoC}{Proof of Concept}
\acrodef{PPO}{Proximal Policy Optimization}
\acrodef{POMDP}{Partially Observable Markov decision process}
\acrodef{PCAP}{Packet Capture}
\acrodef{QoE}{Quality of experience}
\acrodef{QoS}{Quality of Service}
\acrodef{QFI}{Quality of Service Flow Identifier}
\acrodef{QFIs}{Quality of Service Flow Identifiers}
\acrodef{RAM}{Random Access Memory}
\acrodef{RF}{Random Forest}
\acrodef{RL}{Reinforcement Learning}
\acrodef{RMSE}{Root Mean Square Error}
\acrodef{RNN}{Recurrent Neural Network}
\acrodef{RTT}{Round-Trip Time}
\acrodef{RAN}{Radio Access Network}
\acrodef{RTP}{Real-time Transport Protocol}
\acrodef{RIC}{RAN Intelligent Controller}
\acrodef{SDN}{Software-Defined Networking}
\acrodef{SFI2}{Slicing Future Internet Infrastructures}
\acrodef{SLA}{Service-Level Agreement}
\acrodef{SON}{Self-Organizing Network}
\acrodef{SMF}{Session Management Function}
\acrodef{S-NSSAI}{Single Network Slice Selection Assistance Information}
\acrodef{SVM}{Support Vector Machine}
\acrodef{SOPS}{Service-Aware Optimal
 Path Selection}
\acrodef{TQFL}{Trust Deep Q-learning Federated Learning}
\acrodef{TEID}{Tunnel Endpoint Identifier}
\acrodef{TEIDs}{Tunnel Endpoint Identifiers}
\acrodef{UE}{User Equipment}
\acrodef{UEs}{User Equipments}
\acrodef{UPF}{User Plane Function}
\acrodef{UPFs}{User Plane Functions}
\acrodef{PDU}{Packet Data Unit}
\acrodef{URLLC}{Ultra-Reliable and Low Latency Communications}
\acrodef{UAV}{Unmanned Aerial Vehicle}
\acrodef{UAVs}{Unmanned Aerial Vehicles}
\acrodef{UDP}{User Datagram Protocol}
\acrodef{VoD}{Video on Demand}
\acrodef{VR}{Virtual Reality}
\acrodef{AR}{Augmented Reality}
\acrodef{V2V}{Vehicle-to-Vechile}
\acrodef{V2X}{Vehicle-to-Everything}
\acrodef{VNF}{Virtual Network Function}
\acrodef{VNFs}{Virtual Network Functions}

\acrodef{XDP}{eXpress Data Path}

\title{Noisy Neighbor Influence in the Data Plane of Beyond 5G Networks}

\author{\IEEEauthorblockN{Rodrigo Moreira\textsuperscript{1}, Larissa Ferreira {Rodrigues Moreira}\textsuperscript{1}, Tereza C. Carvalho\textsuperscript{2},
Flávio de Oliveira Silva\textsuperscript{3}}
\IEEEauthorblockA{
\textsuperscript{1}Federal University of Viçosa (UFV), Minas Gerais, Brazil\\
\textsuperscript{2}University of São Paulo (USP), São Paulo, Brazil\\
\textsuperscript{3}University of Minho (UMinho), Braga, Portugal\\
Emails: rodrigo@ufv.br, larissa.f.rodrigues@ufv.br, terezacarvalho@usp.br, flavio@di.uminho.pt}
}

\maketitle

\begin{abstract}

Virtualization and containerization enhance the modularity and scalability of mobile network architectures, facilitating customized user services and improving management and orchestration across the network. In the context of the 5th Generation Mobile Network (5G), these advancements contribute to reduced Operational Expenditures (OPEX) and enable sliced-based networking for novel applications and services. However, as beyond fifth-generation (B5G) networks aim to address the remaining challenges regarding network slice isolation, the shared underlying hardware can lead to data plane contention among slices, resulting in the Noisy Neighbor (NN) effect, which may compromise network slicing and Service-Level Agreements (SLAs). We propose a kernel-level instrumentation of the User Plane Function (UPF) to assess the impact of noisy slices on data plane processing. Our findings reveal that even prioritized slices are susceptible to degradation induced by NN, with observable effects on latency metrics pertinent to user experience.

\end{abstract}

\begin{IEEEkeywords}
B5G, Noisy Neighbor, UPF Instrumentation, Slice Isolation, Data-Plane Latency.
\end{IEEEkeywords}

\section{Introduction}\label{sec:introduction}

Cloud computing employs virtualization and containerization to deliver multi-tenant isolation and efficient resource use; the rise of microservices and serverless computing further increases consolidation density, operational agility, and the pace of cloud-native innovation~\cite{zhong_cost-efficient_2020, zhong_machine_2022, copik_software_2024, Golec2024, Andreoli2025}. This paradigm has reshaped mobile network architectures, such as those of \ac{5G} and emerging \ac{B5G}, enabling them to adopt cloud-native designs from the core to the access point and decompose network functions into modular microservices~\cite{Donatti2024}. The result is fine-grained scaling and lower \ac{OPEX}, but also greater susceptibility to cross-tenant performance interference or noise effects ~\cite{Muro2023, Lozano2024}. Even slices engineered with \ac{SLA} guarantees can face resource contention that weakens isolation in practice~\cite{yarkina_multi-tenant_2022, Muro2023, Andreoli2025}.

The \ac{NN} effect occurs when aggressive co-tenants exhaust shared resources and remains a major source of cross-tenant interference in virtualized environments~\cite{Lozano2024}, from \ac{5G} core microservices to virtualized baseband, despite \ac{CPU} pinning, cgroup limits, and cache partitioning. Its impact manifests most intensely in the user plane, where latency spikes and heavy tails erode slice-based differentiation~\cite{luo_optimizing_2024, Volpert2025}. Despite prior work on detection and mitigation in cloud and telecom, there is little kernel-level, per-packet evidence of how \ac{NN}s perturb the mobile data plane, especially within the \ac{UPF} during \ac{GTP-U} decapsulation. Methods based on aggregate throughput, queue counters, or offline profiling miss microbursts and cannot attribute latency to specific kernel stages~\cite{Muro2023, Hong2024, Volpert2025, Andreoli2025}.

We propose kernel instrumentation to measure and attribute the \ac{GTP-U} decapsulation overhead in a container-based \ac{UPF} on Kubernetes. Using bpftrace, an \ac{eBPF}, the probe timestamps at the \ac{GTP} module entry and at \texttt{netif\_rx} are used to compute the per-packet decapsulation latency. It filters a target \ac{UE} \ac{IP} to isolate a sensor flow and tags each sample with a \ac{TEID}/\ac{QFI} context. The result is high-resolution latency traces that reveal how competing flows and co-located workloads modulate the decapsulation cost across traffic classes and slice priorities.

Our main contributions are as follows: (i) we introduce lightweight, kernel-level per-packet instrumentation for the \ac{UPF} \ac{GTP-U} decapsulation that attributes data-plane latency to specific \ac{TEID}/\ac{QFI} flows; (ii) we provide a compact analysis framework with robust summaries and nonparametric tests to quantify the impact of the \ac{NN} on median and tail latencies; (iii) we present empirical evidence that disruptive co-located workloads inflate decapsulation latency even for high-priority \ac{QFI}s, revealing the limits of slice isolation; and (iv) we distill operator guidance for telemetry and orchestration (what to monitor, how to flag early degradation, and how to prioritize mitigation). Our diagnostic framework lays the foundation for runtime mitigation strategies.

The remainder of this paper is organized as follows: section ~\ref{sec:related_work} reviews interference and \ac{NN} literature across cloud and mobile networks. Section~\ref{sec:method} details our proposed method. Section~\ref{sec:evaluation_setup} presents the testbed and evaluation scenarios. Section~\ref{sec:results_and_discussion} presents the results and discussion, and Section~\ref{sec:concluding_remarks} draws conclusions and outlines future work.

\section{Related Work}\label{sec:related_work}

\begin{table*}[t!]
\centering
\caption{Comparison of related work.}
\label{tab:rw-compact}
\resizebox{\textwidth}{!}{
\setlength{\tabcolsep}{3pt} 
\renewcommand{\arraystretch}{1.1} 
\begin{tabular}{lccccccccccc}
\toprule
\textbf{Approach}                                                      & \textbf{5GC} & \textbf{UPF} & \textbf{RAN} & \textbf{NN focus} & \textbf{Online detect.} & \textbf{Mitigation} & \textbf{eBPF} & \textbf{Per-packet (kernel)} & \textbf{K8s-based} & \textbf{Open-src} & \textbf{CP focus} \\ \midrule
Mukute et al.~\cite{Mukute2024}                         & \faCircle          & \faCircleO           & \faCircleO           & \faCircleO                & \faCircleO                      & \faCircleO                  & \faCircle           & \faCircleO                           & \faCircleO                 & \faCircle               & \faCircle               \\
Volpert et al.~\cite{Volpert2025}                           & \faCircleO           & \faCircleO           & \faCircleO           & \faCircle               & \faCircle                     & \faCircleO                  & \faCircle           & \faCircleO                           & \faCircleO                 & \faCircleO      & \faCircleO                \\
Muro et al.~\cite{Muro2023}                                  & \faCircle          & \faCircleO   & \faCircleO           & \faCircle               & \faCircleO                      & \faCircle                 & \faCircleO            & \faCircleO                           & \faCircleO                 & \faCircleO       & \faCircle            \\
Andrade et al.~\cite{Andrade2025}        & \faCircle          & \faCircle          & \faCircleO           & \faCircleO                & \faCircleO                      & \faCircle                 & \faCircleO            & \faCircleO                           & \faCircle                & \faCircle               & \faCircleO                \\
Adeppady et al.~\cite{Adeppady2023}            & \faCircleO           & \faCircleO           & \faCircleO           & \faCircle               & \faCircleO                      & \faCircle                 & \faCircleO            & \faCircleO                           & \faCircle                & \faCircleO       & \faCircleO                \\
Da Silva et al.~\cite{daSilva2025}                 & \faCircle          & \faCircleO           & \faCircleO           & \faCircleO                & \faCircleO                      & \faCircleO                  & \faCircleO            & \faCircleO                           & \faCircle                & \faCircleO       & \faCircle               \\
Lozano et al.~\cite{Lozano2024}                         & \faCircleO           & \faCircleO           & \faCircle          & \faCircle               & \faCircle                     & \faCircle                 & \faCircleO            & \faCircleO                           & \faCircleO                 & \faCircleO       & \faCircleO                \\ \midrule
\textbf{Ours} & \faCircle          & \faCircle          & \faCircleO           & \faCircle               & \faCircleO                      & \faCircleO                  & \faCircle           & \faCircle                          & \faCircle                & \faCircle               & \faCircleO                \\ \bottomrule
\end{tabular}
}
\end{table*}

This section highlights several studies that focused on implementing and examining the impact of simultaneous applications in virtualized or containerized settings. Finally, we present our findings in a comparative table that motivates our instrumentation method.

Mukute et al.~\cite{Mukute2024} provide a control plane centered, feature grounded comparison of open source \ac{5G} cores by mapping network function operations to \ac{5G} procedures and combining macro benchmarking with micro benchmarking using a custom \ac{UE} traffic generator and \ac{eBPF} profiling. They report end to end registration times and per process system call latency and frequency with attribution, show correlation between macro and micro signals that reveals optimization points, and release practical recommendations and replication artifacts.

Volpert et al.~\cite{Volpert2025} study interference that persists under \ac{CPU} cgroup isolation and the limits of detectors that depend on workload profiles or offline analysis. They instrument the Linux scheduler with \ac{eBPF} tracepoints and define two online, workload agnostic metrics, Average Process Scheduling Latency and Average Process Scheduler Preemptions, plus a decision matrix. Across eight contention scenarios on Kubernetes and Argo with stress-ng and nsjail, the metrics track resource isolation and application quality of service, detect noisy neighbors online without per workload models, and separate them from self disruption.

Muro et al.~\cite{Muro2023} build a core focused testbed to quantify noisy neighbor effects in virtualized \ac{5G} cores, emulating a full core and creating a cpulimit based simulator that throttles the LoadCore agent. A factorial campaign with and without a noisy neighbor varies packet rate, packet size, and number of \ac{UE}; they collect throughput, delay, loss, and core \ac{CPU}, train cross validated classifiers, identify packet rate as the dominant driver of core \ac{CPU} demand, and observe throughput reduction with higher delay and loss under noise.

Andrade et al.~\cite{Andrade2025} evaluate slice isolation controls for private \ac{5G} cores on cloud and edge platforms, testing Kubernetes and Linux mechanisms at the \ac{UPF}: \ac{CPU} requests and limits with cgroups, Cilium earliest departure time bandwidth control, and process priority via nice, alone and in combination. In a hospital video conferencing scenario with a prioritized slice under growing background load, 26 experiments with 10 runs and Prometheus telemetry show that \ac{CPU} caps on non priority slices consistently improve the prioritized slice, while bandwidth caps are powerful but can starve others if set too strictly.

Adeppady et al.~\cite{Adeppady2023} design iPlace, a Kubernetes placement heuristic that clusters microservices by contention dissimilarity and enables batch deployment. Using testbed pairwise interference measurements and simulations against several baselines, iPlace approaches optimal solutions, uses between 21 percent and 92 percent fewer servers, and reduces deployment time by 69 percent.

Da Silva et al.~\cite{daSilva2025} pursue low overhead, real time estimation of user experience in container based \ac{5G} cores. Deploying free5GC on Kubernetes and applying stochastic \ac{CPU}, memory, and network stress with Chaos Mesh while issuing continuous registrations, they learn regressors from generic infrastructure metrics to registration latency on a FABRIC testbed with a 24 hour cyclic pattern. \ac{RF} achieves approximately 10.5 percent \ac{MAPE} with a top ten feature subset, which also reduces runtime, outperforming \ac{KNN}, \ac{DT}, \ac{SVM}, and \ac{LSTM} in this setting.

Lozano et al.~\cite{Lozano2024} examine noisy neighbor effects in shared virtual \ac{RAN}. On a Dockerized srsRAN testbed they quantify overhead sources including hyper threading, namespaces, seccomp, context switches, and cache contention, then propose an \ac{O-RAN} compliant non real time \ac{RIC} policy that uses a Relation Network encoder with a \ac{DQN} to select the number of active cores. With two to five virtual base stations under maximum load and five-day traces, they measure throughput, \ac{CPU} assignment, power, instructions per cycle, cache misses, and inference time, identify cache contention as the dominant bottleneck, and show that the policy reaches near oracle performance with up to 30 percent resource savings and about 99.9 percent availability.


Table~\ref{tab:rw-compact} contrasts prior work and ours using binary features, where a filled circle ({\scriptsize \faCircle}) denotes ``Yes'' and a hollow circle ({\scriptsize \faCircleO}) denotes ``No''. The second column (5GC) indicates work targeting the \ac{5G} Core; \ac{UPF} flags user plane processing; RAN marks radio-access analysis; \ac{NN} focus denotes explicit treatment of noisy-neighbor interference; Online detect. indicates run-time detection; Mitigation means the work proposes or evaluates countermeasures; \ac{eBPF} captures the use of kernel observability via \ac{eBPF}/\ac{BPF} toolchains; Per-packet (kernel) means kernel-space, per-packet telemetry/instrumentation; K8s-based signals experimentation/deployment on Kubernetes; Open-src denotes released artifacts enabling reproduction; and Control Plane (CP)  highlights a predominant control-plane orientation (as opposed to data-plane).

\textbf{Positioning.} Prior work spans control plane comparisons, scheduler detectors, slice controls, placement, and user experience prediction at aggregate timescales; we provide kernel level, per packet attribution of \ac{UPF} \ac{GTP-U} decapsulation latency under coexisting \ac{TEID} and \ac{QFI} flows, revealing microbursts and limits of slice isolation that aggregate counters cannot see.

\section{Proposed Method}\label{sec:method}

\begin{figure}[ht]
  \includegraphics[width=0.9\columnwidth]{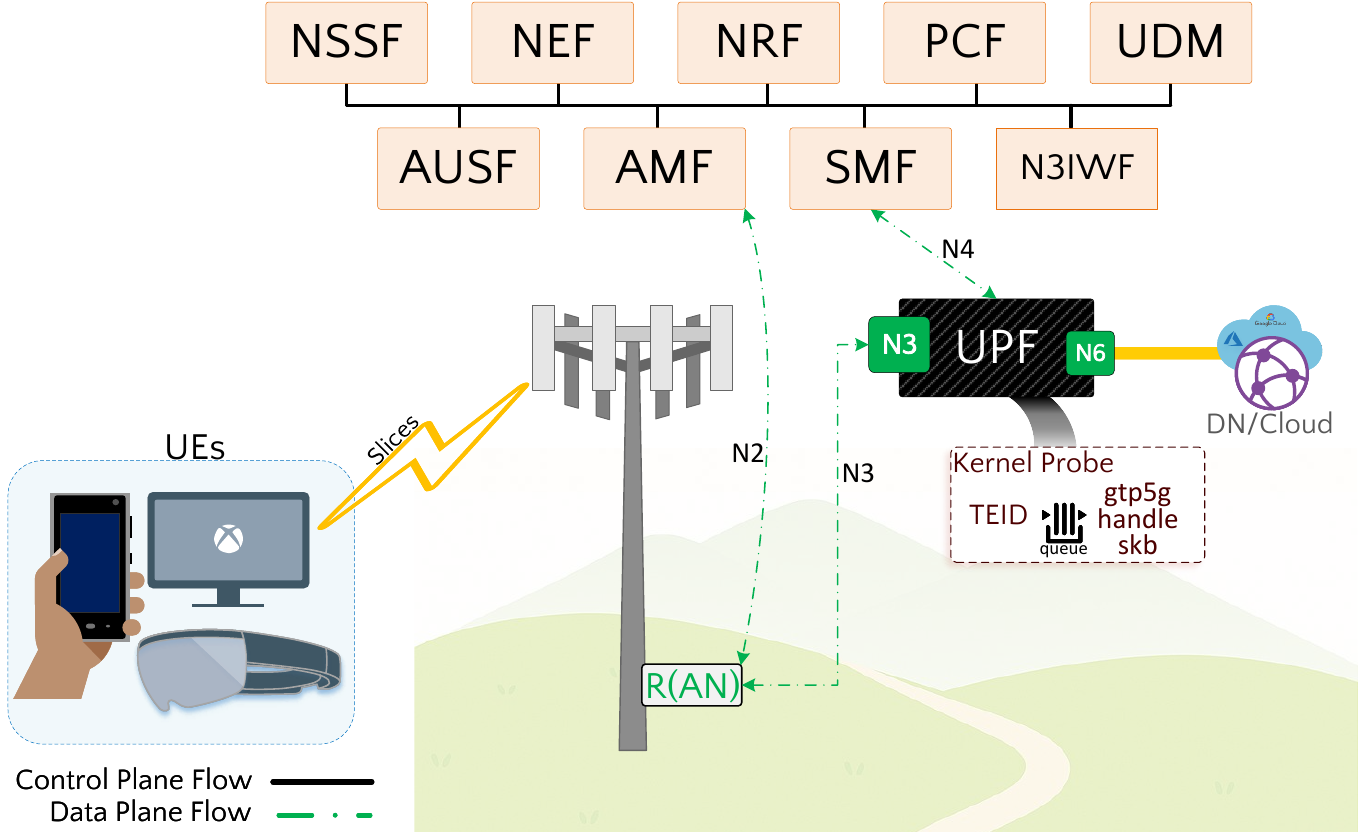}
  \caption{Proposed Instrumentation Method.}
  \label{fig:proposed_method}
\end{figure}

Quantifying the mutual effects of \ac{5G} slices on the data plane is challenging, especially in terms of processing cost and traffic isolation between network slices. We propose a kernel-level bpftrace instrumentation to measure the per-packet overhead of \ac{GTP-U} decapsulation in the \ac{UPF}. As in Fig.~\ref{fig:proposed_method}, we timestamp packets at the entry of \texttt{gtp5g\_handle\_skb\_ipv4} (decapsulation) and again at \texttt{netif\_rx}; their difference is the decapsulation latency. As depicted in Fig.~\ref{fig:proposed_method}, within the \ac{UPF}, the N3 interface serves as the entry point for marking in-kernel packets. Internally, we measured various function callbacks to determine the time elapsed between \ac{GTP} processing and \texttt{netif\_rx} towards N6. The N2 and N4 interfaces constitute the control plane interfaces of the 5G core, whereas N3 functions as the ingress user data interface.


Our proposed probe\footnote{The source code for our proposed approach is accessible at \url{https://github.com/romoreira/5GNF_NoisyNeighbor}.} filters only packets whose inner \ac{UE} \ac{IP} matches a configured sensor \ac{UE} set by the experimenter. On match, it stores \texttt{nsecs} and the \ac{IP} in \ac{BPF} maps keyed by the \texttt{sk\_buff} pointer; when the same \texttt{skb} reaches \texttt{netif\_rx}, it computes $\Delta t$ and emits ``UE IP: $IP_{ue}$ \,|\, Decapsulation Overhead: $\Delta t$ ns,'' then clears the entries. This lightweight design yields high resolution latency for each \ac{TEID} flow and enables a precise evaluation of the \ac{NN} interference in multi-\ac{TEID} environments. The procedure is summarized in Algorithm~\ref{alg:kprobe_algorithm}.

\begin{algorithm}[t!]
\caption{Per packet GTP-U decapsulation overhead via \texttt{bpftrace}}
\label{alg:kprobe_algorithm}
\KwIn{Target UE IP $IP_{\text{target}}$}
\KwOut{Overhead (ns) for matching packets}

\textbf{kprobe} \texttt{gtp5g\_handle\_skb\_ipv4}: parse inner UE IP at \texttt{skb->data+8} as $IP_{ue}$; if $IP_{ue}=IP_{\text{target}}$ then set \texttt{@start[skb]=nsecs}, \texttt{@ip[skb]=$IP_{ue}$}.\\
\textbf{kprobe} \texttt{netif\_rx}: if \texttt{@start[skb]} exists then let $t=\texttt{@start[skb]}$, $IP_{ue}=\texttt{@ip[skb]}$, and $\Delta t=\texttt{nsecs}-t$; print $(IP_{ue},\Delta t)$; delete \texttt{@start[skb]}, \texttt{@ip[skb]}.
\end{algorithm}

\subsection{Influence of Noisy Neighbors on Slice Isolation}\label{subsec:influence_of_noisy}

In \ac{5G}, each \ac{UE} data session forms a \ac{PDU} session with a tunnel identified by a unique \ac{TEID}, allowing the \ac{UPF} to map user traffic to the correct \ac{GTP-U} tunnel. Within a \ac{PDU} session, the \ac{QFI} labels the \ac{QoS} flow and is bound to the \ac{TEID} and \ac{UE}, so each stream receives its intended treatment. In practice, however, our method enables assessment of whether high \ac{QFI} reservations and isolation are guaranteed.

Therefore, we varied the traffic classes and flow characteristics to induce “noisy neighbors” competing slices that create substantial loads or unpredictable bursts, and measured per-packet decapsulation latency across multiple \ac{TEID} flows. This helps us see how well we can keep these flows separate and check if flows with strict \ac{QoS} settings are protected from these noisy neighbors.

\subsection{Statistical Analysis}\label{subsec:statistical_analysis}

To measure the influence inter \ac{5G} slices on container-based environments, we use statistical techniques as follows. For each \ac{UE} traffic pattern and \ac{QFI}, we summarize per-packet decapsulation latency with the median, p90, p95, p99, \ac{IQR}, and \ac{MAD}. Pairwise comparisons against \textit{Baseline} use two-sided Mann–Whitney U tests for median differences. In addition, permutation tests with $10^4$ resamples assess $\Delta$median and $\Delta$p95, and bias-corrected and accelerated (BCa) bootstrap 95\% confidence intervals are reported for both.

Effect sizes are given by Cliff’s $\delta$, 
\[
\delta=\frac{\#(x>y)-\#(x<y)}{n_x n_y},
\]
where $x$ and $y$ are latency samples from the two groups (Baseline as reference vs.\ each alternative pattern: \textit{Anomaly}, \textit{Constant-Rate}, \textit{Multimedia}).

To model distributional effects, we fit quantile regressions at $\tau\in\{0.50,0.95,0.99\}$ (median, high tail, extreme tail):
\begin{multline}
Q_\tau(\Delta t\mid \mathbf{x}) = \beta_0 + \beta_1\,\text{CPU} + \beta_2\,\text{Packets} \\
+ \sum_{q\in\mathcal{Q}}\beta_{3,q}\,\mathbb{I}[\text{QFI}=q]
+ \sum_{c\in\mathcal{C}}\gamma_c\,\mathbb{I}[\text{Traffic}=c]
\end{multline}

where $\Delta t$ is \ac{GTP-U} decapsulation latency in ns, CPU is the instantaneous \ac{UPF} \ac{CPU} utilization (\%), and \texttt{Packets} is the instantaneous packet rate (pps). $\mathcal{Q}$ is the set of observed \ac{QFI}s; $\mathcal{C}=\{Baseline, Anomaly, Constant-Rate, Multimedia\}$. The coefficients are estimated separately for each $\tau$. Quantile regression is preferred to mean-based models because data-plane latencies are skewed and heavy-tailed; upper quantiles (p95, p99) better capture \ac{QoS} behavior.

\section{Evaluation Setup}\label{sec:evaluation_setup}

We ran experiments on the FABRIC testbed~\cite{fabric-2019} using a virtual machine with 32vCPUs and 32GB RAM. The stack was Ubuntu + Kubernetes~v1.28.15, with Free5GC emulating the \ac{5G} core. We monitored system resources with Netdata (network/\ac{I/O}) and \texttt{pidstat} for per–process \ac{CPU}. To avoid bias, the containerized \ac{UPF} was pinned to a single \ac{CPU} core.

\begin{figure}[ht]
  \includegraphics[width=0.85\columnwidth]{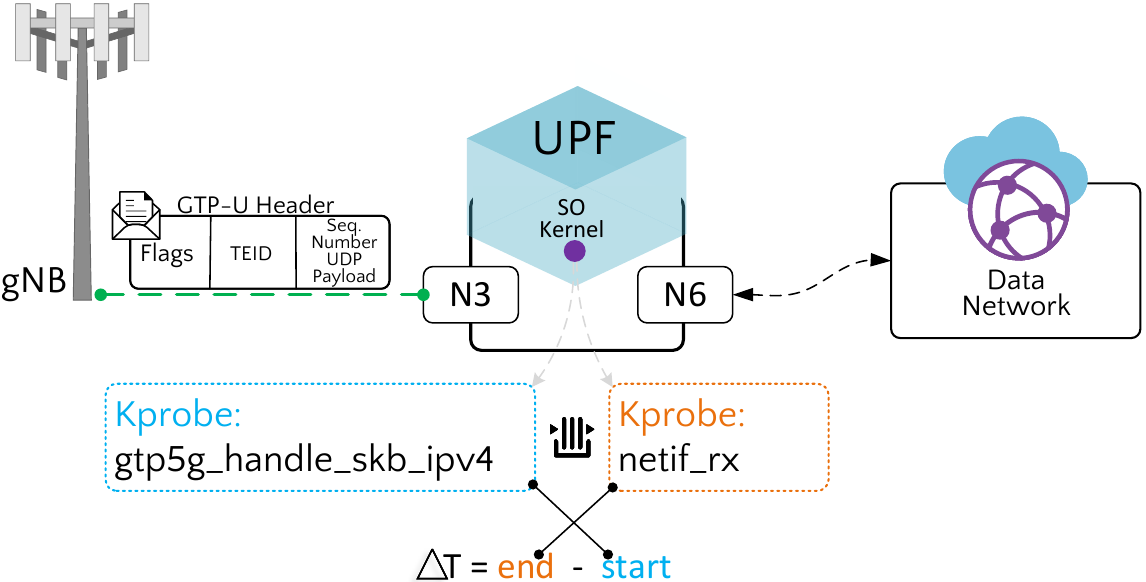}
  \caption{Experimental setup and placement of our instrumentation within the \ac{UPF} data path.}
  \label{fig:experimental_setup}
\end{figure}

As shown in Fig.~\ref{fig:experimental_setup}, the \ac{UPF} was instrumented with \ac{eBPF} probes to measure per–packet latency from ingress at the N3 interface to post–decapsulation delivery on eth0 (N6). Traffic was generated by a \ac{5G} traffic tool emitting \ac{GTP-U} packets with valid \ac{TEID}s and \ac{QFI}s across distinct patterns: \ac{CBR} and \ac{PCAP} replays of multimedia applications (Table~\ref{tab:pcap_summary}). Future patterns such as \ac{URLLC} and massive \ac{IoT} are out of this scope. All replay traces were 100\% \ac{UDP} with no \ac{RTP} streams detected.

\begin{table}[ht]
\caption{Replayed multimedia PCAPs (Data in MiB; Avg in Mb/s).}
\label{tab:pcap_summary}
\centering
\setlength{\tabcolsep}{2.1pt}
\renewcommand{\arraystretch}{1.1} 
\begin{tabular}{lcccccc}
\toprule
\textbf{Application} & \textbf{Dur (s)} & \textbf{Pkts (k)} & \textbf{Data (MiB)} & \textbf{Avg (Mb/s)} & \textbf{Pps} & \textbf{Mean (B)}\\
\midrule
Facebook  & 43.8 & 13.2 & 14.7 & 2.8 & 300.5 & 1171 \\
Instagram & 66.1 & 27.1 & 30.9 & 3.9 & 410.8 & 1193 \\
LinkedIn  & 70.1 & 12.5 & 11.1 & 1.3 & 178.1 & 929 \\
PS Now    & 59.4 & 17.5 & 15.1 & 2.1 & 294.2 & 906 \\
Spotify   & 146.3 & 34.9 & 28.1 & 1.6 & 238.8 & 842 \\
TikTok    & 63.9 & 40.7 & 37.8 & 5.0 & 638.0 & 974 \\
Twitter   & 68.9 & 13.0 & 12.0 & 1.5 & 188.4 & 967 \\
Wikipedia & 77.4 & 4.8  & 4.1  & 0.4 & 62.3  & 883 \\
YouTube   & 63.4 & 14.2 & 16.6 & 2.2 & 224.0 & 1226 \\
\bottomrule
\end{tabular}
\vspace{-0.6em}
\end{table}

We established \ac{PDU} sessions in advance to ensure proper \ac{UPF} processing. \ac{NN} scenarios were created by co–locating additional active \ac{PDU} sessions on the same \ac{UPF} to stress its \ac{CPU} and \ac{I/O}, while monitoring by a \ac{PDU} sensor. Two scenarios were evaluated: (i) a sensor \ac{UE} with a \ac{TEID} mapped to a high–priority \ac{QFI} while replaying diverse neighbor traffic (Table~\ref{tab:lat_summary_traffic}) and (ii) the same sensor \ac{UE} bound to a low–priority \ac{QFI}. Each neighbor had a valid \ac{TEID} and active \ac{PDU} session, forcing the traversal of \texttt{gtp5g\_handle\_skb\_ipv4}. The sensor \ac{UE} sent fixed–rate \ac{ICMP} probes, and we correlated timestamped per–packet latencies with concurrent compute and network metrics to assess the impact of different neighbor priorities.

\section{Results and Discussion}\label{sec:results_and_discussion}

We evaluated our method using a real \ac{5G} testbed core and instrumented the \ac{UPF} with our per–packet probe from Section~\ref{sec:method}. We analyzed decapsulation latency across traffic classes and \ac{QoS} flows, tested differences against the \textit{Baseline}, and modeled distributional effects with quantile regression. 

\subsection{Descriptive Analysis}

Table~\ref{tab:lat_summary_traffic} summarizes \ac{UPF} latency by \ac{UE} traffic. Baseline (\ac{ICMP}) has median $16.33$ns (p95 $28.23$ns). Anomaly is similar in center/tail but more dispersed (IQR $10.68$ns vs.\ $4.19$ns). Constant-Rate \ac{CBR} is lower (median $12.75$ns; p95 $23.62$ns), while Multimedia lies between them (median $14.88$ns; p95 $19.32$ns). The \ac{QFI} breakdown (Table~\ref{tab:lat_summary_qfi}) shows heterogeneity: QFI~9 has a slightly higher median than QFI~1 but a lower p95 value.

\begin{table}[!htbp]
\centering
\caption{Latency summary by UE Traffic class. Values in ns.}
\label{tab:lat_summary_traffic}
\renewcommand{\arraystretch}{1.1} 
\setlength{\tabcolsep}{7pt}
\begin{tabular}{lrrrrrr}
\toprule
\textbf{UE Traffic} & \textbf{n} & \textbf{Median} & \textbf{p90} & \textbf{p95} & \textbf{p99} & \textbf{IQR} \\
\midrule
Anomaly & 60 & 16.14 & 26.33 & 28.62 & 38.46 & 10.68 \\
Baseline & 60 & 16.33 & 23.33 & 28.23 & 31.35 & 4.19 \\
Constant-Rate & 60 & 12.75 & 17.97 & 23.62 & 27.99 & 4.01 \\
Multimedia & 60 & 14.88 & 18.93 & 19.32 & 22.90 & 4.74 \\
\bottomrule
\end{tabular}
\end{table}

The \ac{QFI}-level breakdown (Table~\ref{tab:lat_summary_qfi}) reveals that \ac{QFI}~9 presents a slightly higher median latency than \ac{QFI}~1, but a lower p95 value, suggesting heterogeneous effects worth further modeling.

\begin{table}[!htbp]
\centering
\caption{Latency summary by QFI. Values in ns.}
\label{tab:lat_summary_qfi}
\setlength{\tabcolsep}{9pt}
\begin{tabular}{rrrrrrr}
\toprule
\textbf{QFI} & \textbf{n} & \textbf{Median} & \textbf{p90} & \textbf{p95} & \textbf{p99} & \textbf{IQR} \\
\midrule
1 & 120 & 14.35 & 25.11 & 27.49 & 31.88 & 6.59 \\
9 & 120 & 15.27 & 19.28 & 22.66 & 33.09 & 5.24 \\
\bottomrule
\end{tabular}
\end{table}

\subsection{Group Comparisons}

Here, we report on whether traffic patterns materially shifted user-plane latency and thus affected slice isolation, and we compared each class to the baseline. Permutation tests ($10^4$ resamples), Mann–Whitney U, and Cliff’s $\delta$ (Table~\ref{tab:group_tests}) indicate: \textit{Anomaly} vs.\ \textit{Baseline} with no significant change. \textit{Constant-Rate} reduces median by $-3.58$ns ($p<10^{-3}$; $\delta=-0.48$, large). \textit{Multimedia} yields a moderate median reduction ($-1.45$ns, $p\approx0.038$) and a significant p95 decrease ($-8.91$ns, $p\approx0.014$; $\delta=-0.27$, small). Detailed intervals (Table~\ref{tab:group_tests}): \textit{Constant-Rate} $\Delta$median $-3.576$ns [95\% CI $-4.950,-2.299$]; \textit{Multimedia} $\Delta$median $-1.452$ns [$-2.772,-0.475$], $\Delta$p95 $-8.907$ns [$-11.752,-1.519$]. \textit{Anomaly} effects are negligible.

\begin{table}[ht]
\centering
\caption{Baseline vs. other traffic classes: effect sizes and statistical tests.}
\label{tab:group_tests}
\resizebox{\columnwidth}{!}{%
\setlength{\tabcolsep}{3pt}
\renewcommand{\arraystretch}{1.1} 
\begin{tabular}{lrrrrrrl}
\toprule
\textbf{Comparison} & \textbf{$\Delta$Median} & \textbf{$p_{\text{perm}}$} & \textbf{$\Delta$p95} & \textbf{$p_{\text{perm}}$} & \textbf{MW $p$} & \textbf{$\delta$} & \textbf{Magnitude} \\
\midrule
Anomaly & -0.19 & 0.91 & 0.39 & 0.87 & 0.688 & 0.043 & negligible \\
Const.-Rate & -3.58 & \textbf{0.0002} & -4.61 & 0.27 & \textbf{$5\times 10^{-6}$} & -0.484 & large \\
Multimedia & -1.45 & \textbf{0.038} & -8.91 & \textbf{0.014} & 0.011 & -0.269 & small \\
\bottomrule
\end{tabular}%
}
\end{table}

\begin{figure*}[ht!]
    \centering

    \begin{subfigure}[b]{0.31\textwidth}
        \centering
        \includegraphics[width=\textwidth]{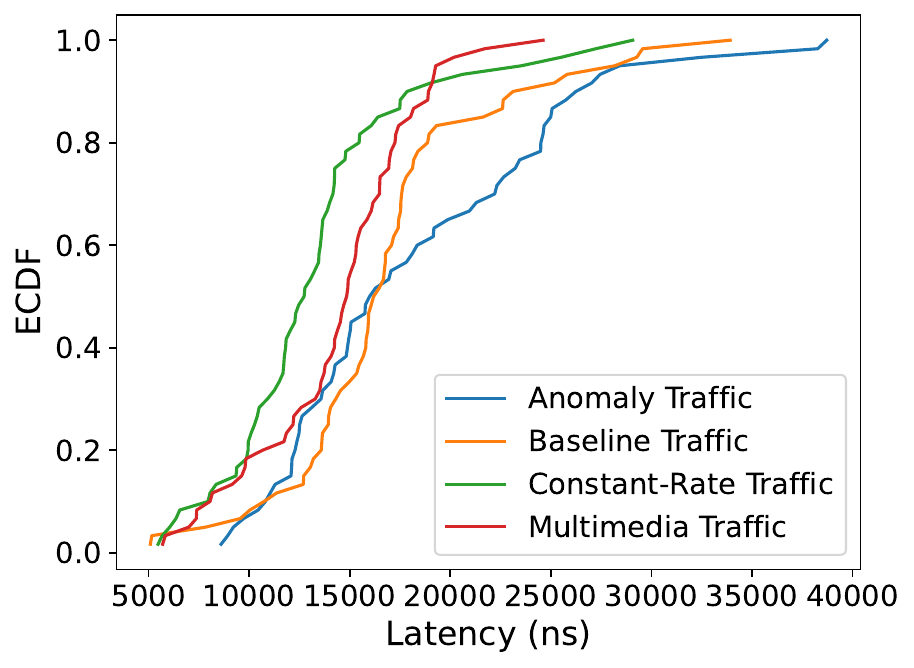}
        \caption{ECDF of \ac{UPF} Processing Latency.}
        \label{fig:ecdf_latency}
    \end{subfigure}
    \hfill
    \begin{subfigure}[b]{0.285\textwidth}
        \centering
        \includegraphics[width=\textwidth]{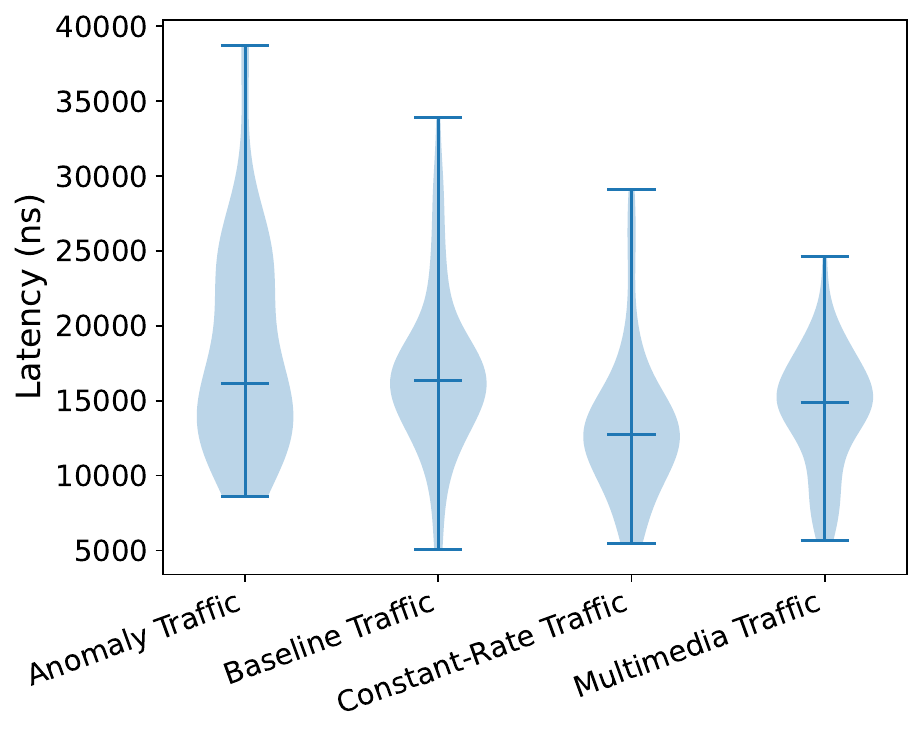}
        \caption{Latency distribution (violin plots).}
        \label{fig:violin_latency}
    \end{subfigure}
    \hfill
    \begin{subfigure}[b]{0.31\textwidth}
        \centering
        \includegraphics[width=\textwidth]{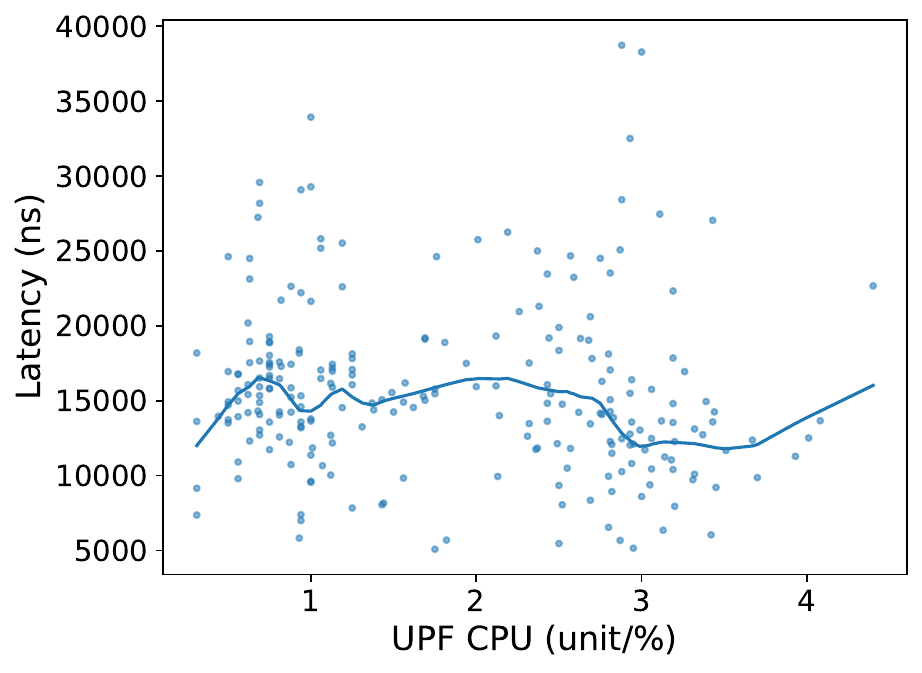}
        \caption{Latency vs. \ac{CPU} utilization.}
        \label{fig:scatter_cpu_latency}
    \end{subfigure}

    \caption{Latency characteristics across traffic classes and CPU load.}
    \label{fig:latency_comparison}
\end{figure*}

\subsection{Quantile Regression Analysis}

To reveal which factors drive typical and tail latency, we model the full conditional distribution via quantile regression at $\tau!\in!{0.50,0.95,0.99}$ using \ac{CPU}, packet rate, \ac{QFI}, and traffic class as predictors. Quantile models at $\tau\in\{0.50,0.95,0.99\}$ (Table~\ref{tab:quantile_reg}) include \ac{UPF} CPU, packet rate, \ac{QFI}, and traffic dummies. CPU had a negative association with latency at p50 ($-1.28$ns, $p\approx0.016$) and p95 ($-4.10$ns, $p\approx0.038$), indicating better tail behavior with more \ac{CPU}, whereas packets were not significant. QFI~9 is marginal at p50. \textit{Constant-Rate} and \textit{Multimedia} coefficients are strongly negative at p50/p95, aligning with non–parametric tests. p99 shows convergence limits (extreme–tail sparsity); increasing \texttt{max\_iter} or simplifying the model mitigated this.

\begingroup
\setlength{\tabcolsep}{4pt}        
\renewcommand{\arraystretch}{1.1} 
\footnotesize                     
\begin{table}[!htbp]
\centering
\caption{Quantile regression coefficients (ns) and $p$-values for selected predictors.}
\label{tab:quantile_reg}
\begin{tabular}{l c c c c c}
\toprule
\textbf{Quantile} & $n$ & CPU & Packets & QFI$_9$ & Const.-Rate \\
\midrule
p50 & 240 &
  \makecell{$\beta=-1.28$\\$p=0.016$} &
  \makecell{$\beta=0.86$\\$p=0.361$} &
  \makecell{$\beta=1.16$\\$p=0.066$} &
  \makecell{$\beta=-3.08$\\$p<0.001$} \\ \hline
p95 & 240 &
  \makecell{$\beta=-4.10$\\$p=0.038$} &
  \makecell{$\beta=4.69$\\$p=0.120$} &
  \makecell{$\beta=0.56$\\$p=0.740$} &
  \makecell{$\beta=-10.31$\\$p<0.001$} \\ \hline
p99 & 240 &
  \makecell{$\beta=-4.77$\\$p=\text{--}$} &
  \makecell{$\beta=2.84$\\$p=\text{--}$} &
  \makecell{$\beta=-0.12$\\$p=\text{--}$} &
  \makecell{$\beta=-19.02$\\$p=\text{--}$} \\
\bottomrule
\end{tabular}
\end{table}
\endgroup

\subsection{Visualization Analysis}

\ac{ECDFs} (Fig.~\ref{fig:ecdf_latency}) show the tightest distribution for \textit{Constant-Rate}, followed by \textit{Multimedia}; \textit{Anomaly} tracks \textit{Baseline} with a slightly heavier tail, matching Table~\ref{tab:lat_summary_traffic}. Violin plots (Fig.~\ref{fig:violin_latency}) confirm the asymmetry and long upper tails, which are most visible for bursty traffic. The CPU–latency scatter with \ac{LOWESS} (Fig. ~\ref{fig:scatter_cpu_latency}) exhibits a downward trend and a much larger variance at low CPU, consistent with the negative CPU coefficients.

\subsection{Discussion and Insights}

The results support the hypothesis that traffic patterns significantly influence \ac{UPF} processing latency, particularly in the upper quantiles. 

\textbf{Predictability improves latency.} Constant-rate traffic not only lowers the median but also shows the largest tail latency improvements, likely owing to its predictable nature, which enables better \ac{CPU} scheduling and queue management. The \ac{ECDF} (Fig.\ref{fig:ecdf_latency}) and violin plots (Fig.~\ref{fig:violin_latency}) further illustrates this effect, with Constant-Rate exhibiting a narrow, steep distribution and minimal upper-tail elongation.

\textbf{Moderate traffic, moderate gains.} Multimedia traffic also benefits from reduced latency, although to a lesser extent. The violin plot (Fig.\ref{fig:violin_latency}) reveals a moderate spread with a less pronounced tail than Baseline, confirming its intermediate position between highly predictable and bursty patterns. The negligible effect of Anomaly Traffic is consistent across the \ac{ECDF} (Fig.~\ref{fig:ecdf_latency}) and violin plots (Fig.~\ref{fig:violin_latency}) visualizations, where its curve and density shape remain close to the Baseline but with a slightly heavier tail, suggesting that the current resource isolation and scheduling mechanisms can absorb bursty interference without major degradation in the data plane.

\textbf{Resource allocation matters.} The negative \ac{CPU} coefficients in the quantile models highlight the importance of adequate \ac{CPU} allocation for maintaining low-tail latencies. This relationship is visually reinforced in the CPU–latency scatter plot (Fig.~\ref{fig:scatter_cpu_latency}), where higher \ac{CPU} availability is associated with consistently lower latencies, and low-CPU regions show both higher medians and markedly larger variances. The differences between the \ac{QFI} values indicate that the \ac{QoS} flows experience heterogeneous processing performance, which warrants further analysis of the interaction effects (CPU$\times$QFI, Packets$\times$QFI) in future studies.

\textbf{Stable flows reduce latency.} Relative to Baseline traffic, Constant-Rate lowers the median by $-3.576$\,ns (95\% CI $[-4.950,-2.299]$\,ns; $p_{\text{perm}}=2\times10^{-4}$; Cliff’s $\delta=-0.484$, large), while Multimedia reduces the median by $-1.452$\,ns (95\% CI $[-2.772,-0.475]$\,ns; $p_{\text{perm}}\approx0.038$) and p95 by $-8.907$\,ns (95\% CI $[-11.752,-1.519]$\,ns; $p_{\text{perm}}\approx0.014$); see Table~\ref{tab:group_tests}. These gains align with narrower IQRs for Constant-Rate and Multimedia (4.010\,ns and 4.736\,ns) versus Anomaly (10.684\,ns) in Table~\ref{tab:lat_summary_traffic}, and with the steeper \ac{ECDF} and slimmer violins in Figs.~\ref{fig:ecdf_latency}–\ref{fig:violin_latency}. Quantile regression (Table~\ref{tab:quantile_reg}) shows negative \ac{CPU} coefficients at p50 and p95 (e.g., p95: $-4.10$\,ns, $p\approx0.038$), consistent with the downward trend and higher variance at low \ac{CPU} values in Fig.~\ref{fig:scatter_cpu_latency}. Instability at p99 reflects rare but large outliers, rather than a systematic effect.

\textbf{Consistency.} Figs.~\ref{fig:ecdf_latency}–\ref{fig:scatter_cpu_latency} confirm the statistics: the \ac{ECDF} for Constant-Rate rises sharply (tight spread), while Anomaly’s heavier tail matches its larger IQR in Table~\ref{tab:lat_summary_traffic}. Violin plots show a slim constant rate and a moderately wider multimedia, consistent with lower dispersion. The CPU–latency scatter concentrates the highest delays and variance at low CPU, aligning with the negative \ac{CPU} coefficients and underscoring the need for resource headroom to control the tails.

\section{Concluding Remarks}\label{sec:concluding_remarks}

We presented a kernel level, per packet instrumentation for a container based \ac{UPF} that attributes \ac{GTP-U} decapsulation latency under \ac{NN} conditions. Quantile regression shows that \ac{UPF} \ac{CPU} is the main lever for controlling latency tails, improving both typical and worst case delay. Heterogeneity across \ac{QFI}s persists, and extreme tail inflation under contention reveals practical limits of slice isolation. Operators may keep \ac{CPU} headroom at the \ac{UPF}, favor rate stable shaping for latency sensitive flows, and monitor the \ac{UPF} path with lightweight \ac{eBPF} rather than only aggregate counters.

In future work, we will close the loop between per-packet telemetry and control for online \ac{NN} mitigation, study interaction effects, extend the method to highly programmable dataplanes, and cover memory and network devices effects and \ac{eBPF} overhead. Promising directions include telemetry-driven control that preserves the \ac{CPU} headroom, priority-aware pacing, and earliest departure time scheduling to bound queuing for critical \ac{QFI}s. Our findings pave the way for exploring high-fidelity isolation mechanisms for \ac{B5G} networks.

\section*{Acknowledgment}

The authors thank the FAPEMIG (Grant \#APQ00923-24), FAPESP MCTIC/CGI Research project 2018/23097-3 - SFI2 - Slicing Future Internet Infrastructures. FCT has also supported this work – Fundação para a Ciência e Tecnologia within the R\&D Unit Project Scope UID/00319/Centro ALGORITMI.

\bibliographystyle{IEEEtran}
\bibliography{references}

\end{document}